\documentclass[twocolumn, 10pt, aps, prb, superscriptaddress]{revtex4-1}

\usepackage{amsmath,amssymb,mathtools,xcolor,bbold}
\usepackage{graphicx}
\usepackage[T1]{fontenc}
\usepackage{hyperref}
\usepackage[utf8]{inputenc}
\usepackage{color}
\usepackage{soul}
\usepackage{ulem}

\newcommand{\Dlog}{{\rm Dlog}}
\def\rmi{{\rm {i}}}

\inputencoding{utf8}

\begin{document}

\title{Linked cluster expansions for open quantum systems on a lattice}

\author{Alberto Biella}
\affiliation{Universit\'{e} Paris Diderot, Sorbonne Paris Cit\'{e}, Laboratoire Mat\'{e}riaux et Ph\'{e}nom\`{e}nes Quantiques, CNRS-UMR7162, 75013 Paris, France}

\author{Jiasen Jin}
\affiliation{School of Physics, Dalian University of Technology, 116024 Dalian, China}

\author{Oscar Viyuela}
\affiliation{Department of Physics, Massachusetts Institute of Technology, Cambridge, MA 02139, USA}
\affiliation{Department of Physics, Harvard University, Cambridge, MA 02318, USA}

\author{Cristiano Ciuti}
\affiliation{Universit\'{e} Paris Diderot, Sorbonne Paris Cit\'{e}, Laboratoire Mat\'{e}riaux et Ph\'{e}nom\`{e}nes Quantiques, CNRS-UMR7162, 75013 Paris, France}
  
\author{Rosario Fazio}
\affiliation{ICTP, Strada Costiera 11, I-34151 Trieste, Italy}
\affiliation{NEST, Scuola Normale Superiore and Istituto Nanoscienze-CNR, I-56126 Pisa, Italy}

\author{Davide Rossini}
\affiliation{Dipartimento di Fisica, Universit\`a di Pisa and INFN, Largo Pontecorvo 3, I-56127 Pisa, Italy}

\date{\today}

\begin{abstract}
  We propose a generalization of the linked-cluster expansions to study driven-dissipative
  quantum lattice models, directly accessing the thermodynamic limit of the system.
  Our method leads to the evaluation of the desired extensive property onto small connected
  clusters of a given size and topology.
  We first test this approach on the isotropic spin-1/2 Hamiltonian in two dimensions,
  where each spin is coupled to an independent environment that induces incoherent spin flips.
  Then we apply it to the study of an anisotropic model displaying
  a dissipative phase transition from a magnetically ordered to a disordered phase.
  By means of a Pad\'e analysis on the series expansions for the average magnetization,
  we provide a viable route to locate the phase transition and to extrapolate
  the critical exponent for the magnetic susceptibility.
\end{abstract}

\maketitle

\section{Introduction}

The recent technological breakthroughs in the manipulation of many-body systems coupled to an external bath
are setting the ground for a careful testing of a new wealth
of physical phenomena in the quantum realm~\cite{kasprzak2006, syassen2008, baumann2010}.
Specifically, several promising experimental platforms aimed at investigating the scenario emerging
from driven-dissipative quantum many-body systems have been recently proposed and realized in the lab.
The most remarkable ones are atomic and molecular optical systems through the use of Rydberg atoms, trapped
ions or atomic ensembles coupled to a condensate reservoir~\cite{Muller_2012},
arrays of coupled QED cavities~\cite{Houck_2012}, or coupled optomechanical resonators~\cite{Ludwig_2013}.
These implementations are scalable enough to enable the construction of tunable
and interacting artificial lattice structures with hundreds of sites.

The coupling between different unit cells can give rise to a plethora of cooperative phenomena determined
by the interplay of on-site interactions, nonlocal (typically nearest-neighbor) processes,
and dissipation~\cite{Tomadin_rev, Hartmann_rev, Sieberer_2016, LeHur_rev, Angelakis_rev}.
Recently, a large body of theoretical works has been devoted to the investigation of the collective behavior
emerging in dynamical response~\cite{Tomadin2010}, many-body spectroscopy~\cite{Carusotto_2009, Grujic_2012, Rivas_2014},
transport~\cite{biella2015, angelakis2015, mertz2016, savona2017_01, savona2017_02}, 
as well as stationary properties.
In the latter context, a careful engineering of the coupling between the system and the environment
can stabilize interesting many-body phases in the steady state~\cite{Diehl2008, Verstraete2009}.
The phase-diagram of such lattice systems has been predicted to be incredibly
rich~\cite{hartmann2010, umucalilar2012, jin2013, Yuge_2014, hoening2014, chan2015, wilson2016, ff2017} and can display
spontaneous ordering associated with the breaking of a discrete~\cite{Lee_2011, Lee_2013, savona2017}
or continuous symmetry~\cite{jose2017, biella2017} possessed by the model.
Recently, the critical behavior emerging at the onset of phase transitions started to be investigated
by means of different analytical and numerical approaches~\cite{torre2012, sieberer2013, marino2016, Rota_2017}.

Theoretically, while at equilibrium we have reached a fairly good understanding
of several aspects of the many-body problem under the framework of textbook statistical mechanics,
this is no longer the case for quantum systems coupled to some external bath.
In such case, we are indeed facing an inherently out-of-equilibrium situation, where the Hamiltonian
of the system $\hat H$ is no longer capable to describe it in its whole complexity,
and the environmental coupling needs to be accounted for and suitably modeled.
Due to the intrinsic difficulty of the problem, a number of approximations are usually considered,
which assume a weak system-bath coupling, neglect memory effects in the bath, and discard fast oscillating terms.
In most of the experimental situations with photonic lattices, these assumptions
are typically met~\cite{Houck_2012, Fitzpatrick_2017}.

As a result, in many cases of relevance, the coupling to the environment leads to a Markovian dynamics of the system's
density matrix $\rho$, according to a master equation in the Lindblad form~\cite{Petruccione_book}:
\begin{equation}
  \partial_t \rho = \mathbb{L} [\rho] = - \rmi [ \hat H,\rho] + \mathbb{D}[\rho],
  \label{eq:Master}
\end{equation}
where $\mathbb{L}$ denotes the so called Liouvillian superoperator (we will work in units of $\hbar = 1$).
While the commutator in the r.h.s.~of Eq.~\eqref{eq:Master} accounts for the unitary part of the dynamics, the dissipative processes are ruled by 
\begin{equation}
  \mathbb{D}[\rho] = \sum_j \Big[ \hat L_j \rho \hat L_j^\dagger - \tfrac12 \big\{ \hat L_j^\dagger \hat L_j , \rho \big\} \Big],
\end{equation}
where $\hat L_j$ are suitable local jump operators that describe the incoherent coupling to the environment.
The master equation~\eqref{eq:Master} covers a pivotal role in the treatment of open quantum systems,
since it represents the most general completely-positive trace preserving dynamical semigroup~\cite{Rivas_book}.
In the following we will restrict our attention to it, and specifically address
the steady-state (long-time limit) solution $\rho_{\rm SS} = \lim_{t \to \infty} \exp(\mathbb{L}t) \rho(0)$
(and thus $\partial_t\rho_{\rm SS} =0$) in situations where the steady state is guaranteed to be unique~\cite{albert2014}. 

Solving the long-time dynamics ruled by Eq.~\eqref{eq:Master} for a many-body system is a formidable,
yet important, task.
Indeed contrary to equilibrium situations, the effect of short-range correlations can be dramatic in a driven-dissipative context, 
and thus they deserve an accurate treatment through the (in principle) full many-body problem. 
Exact solutions are restricted to very limited classes of systems,
which are typically represented by quadratic forms in the field operators and specific jump terms~\cite{Prosen_2008}.
A number of viable routes have been thus proposed, in the recent few years.
Under certain hypotheses, analytic approaches such as perturbation theory~\cite{Li_2016} or renormalization-group techniques
based on the Keldysh formalism~\cite{Sieberer_2016, Maghrebi2015} are possible.
However, their limited regime of validity calls for more general numerical methods which do not suffer these limitations.

From a computational point of view, the main difficulty resides in the exponential growth of the many-body
Hilbert space with the number $N$ of lattice sites. Moreover, the non-Hermitian Liouvillian superoperator $\mathbb{L}$
acts on the space of density matrices (whose dimension is the square of the corresponding Hilbert space dimension),
and its spectral properties are generally much more difficult to be addressed
than the low-lying eigenstates of a Hamiltonian system.
The difficulty remains even for the fixed point of the dynamics $\rho_{\rm SS}$,
that is the density matrix associated with the zero-eigenvalue of $\mathbb{L}$.

While in one dimension tensor-network approaches based on a straightforward generalization of
matrix product states to operators can be effective~\cite{Verstraete_2004, Zwolak_2004, Prosen_2009}
and alternative strategies have been proposed in order to improve
their performances~\cite{Cui_2015, Mascarenhas_2015, Werner_2016},
going to higher dimensions is much harder.
Numerical strategies specifically suited for this purpose have been recently put forward,
including cluster mean-field~\cite{Jin_2016},
correlated variational Ans\"atze~\cite{Degenfeld_2014, Weimer_2015},
truncated correlation hierarchy schemes~\cite{Casteels_2016},
corner-space renormalization methods~\cite{Finazzi_2015}, and even two-dimensional tensor-network structures~\cite{Orus_2016}.
The nonequilibrium extension of the dynamical mean-field theory (which works directly in the thermodynamic limit) 
has been also proved to be very effective in a wide class of lattice systems~\cite{tsuji2009,amaricci2012,aoki2014}.
Each of such methods presents advantages and limitations, and typically performs better on specific regimes.

In this paper we will adapt a class of techniques that, in the past, has revealed to be extremely useful
and versatile in the study of thermal and quantum phase transitions~\cite{Oitmaa_book}.
The key idea consists in computing extensive properties of lattice systems in the thermodynamic limit,
out of certain numerical series expansions. The method, dubbed linked-cluster expansion (LCE),
sums over different contributions associated to clusters of physical sites.
In combination with perturbation theories, LCEs have already proved their worth
in the context of equilibrium statistical mechanics,
both in classical and quantum systems (see Ref.~\onlinecite{Oitmaa_book} and references therein).
Their predictive power lies beyond the range of validity of the perturbation expansion:
using established tools for the analysis of truncated series~\cite{Yang_1952}, it has been possible
to study equilibrium quantum phase transitions, and extract critical exponents.
Here we focus on numerical linked-cluster expansions (NLCEs), where the $k$-th order contribution in the LCE
is obtained by means of exact diagonalization techniques on finite-size clusters with $k$ sites~\cite{Rigol_2006}.
The NLCE has been successfully employed in order to evaluate static properties at zero and finite
temperature~\cite{Rigol_2007}, as well as to study the long-time dynamics and thermalization
in out-of-equilibrium closed systems~\cite{Rigol_2014, Mallayya_2017}.
Moreover it has also revealed its flexibility in combination with other numerical methods that can be used to
address finite-size clusters, such as density-matrix renormalization group algorithms~\cite{Bruognolo_2017}.
Nonetheless, to the best of our knowledge, it has never been applied in the context of open quantum systems.

Here we see NLCE at work in an interacting two-dimensional spin-1/2 model with incoherent spin relaxation~\cite{Lee_2013},
which is believed to exhibit a rich phase diagram, and represents a testing ground
for strongly correlated open quantum systems~\cite{Jin_2016, Rota_2017, Orus_2016}.
We will test our method both far from critical points, and in the proximity of a phase transition:
in the first case NLCE allows us to accurately compute the value of the magnetization, while in the latter
we are able to estimate the critical point as well as the critical exponent $\gamma$
for the divergent susceptibility.

The paper is organized as follows.
In Sec.~\ref{sec:method} we introduce our NLCE method and discuss how it can be applied
to the study of the steady-state of a Markovian Lindblad master equation.
The NLCE is then benchmarked in a dissipative two-dimensional spin-1/2 XYZ model (Sec.~\ref{sec:Model}).
By properly tuning the coupling constants of the Hamiltonian, we are able to
study steady-state properties far away from any phase boundary (Sec.~\ref{sec:XXX}),
and a more interesting scenario exhibiting a quantum phase transition from a paramagnetic
to a ferromagnetic phase (Sec.~\ref{sec:XYZ}). 
In the latter case we discuss a simple strategy (based on the Pad\'e analysis of the expansion)
in order to locate the critical point and to extrapolate the critical exponent $\gamma$.
Finally, Sec.~\ref{sec:concl} is devoted to the conclusions.

\section{Linked-cluster method}
\label{sec:method}

We start with a presentation of the NLCE formalism~\cite{Rigol_2006}, unveiling its natural applicability
to the study of driven-dissipative quantum systems whose dynamics is governed by a Lindblad master equation.
We follow an approach that is routinely employed in series expansions
for lattice models, such as high-temperature classical expansions~\cite{Oitmaa_book}.
Since we are interested in the steady-state properties of the system, our target
objects will be the expectation values of generic extensive observables $\hat {\cal O}$ onto
the asymptotic long-time limit solution $\rho_{\rm SS}$ of the master equation: 
${\cal O} = {\rm Tr} \big[\hat {\cal O} \rho_{\rm SS} \big]$.
In practice, for each cluster appearing in the expansion, the steady-state density matrix $\rho_{\rm SS}$ is reached 
by time-evolving a random initial state according to the master equation~\eqref{eq:Master} by means of a fourth-order
Runge-Kutta method.
We stress that there are no restrictions in the limits of applicability of
this approach to different scenarios for homogenous systems, which can be straightforwardly extended to the case
of generic non-Markovian master equations and/or non-equilibrium states $\rho(t)$.
Therefore, boundary-driven systems~\cite{biella2015, mertz2016, savona2017_01, savona2017_02, buca2017}
and disordered lattices~\cite{biondi2015} do not fit within this framework.

Let us first write the Liouvillian operator ${\mathbb L}$ as a sum of local terms ${\mathbb L}_k$,
each of them supposedly acting on few neighbouring sites.
For the sake of simplicity and without loss of generality,
each term ${\mathbb L}_k$ only couples two neighboring sites:
\begin{equation}
  {\mathbb L} = \sum_{k} \alpha_k {\mathbb L}_k = \sum_{\langle i,j \rangle} \alpha_{ij} {\mathbb L}_{ij} , 
\end{equation}
where $\alpha_{ij}$ denotes the local coupling strength, and the index $k=(i,j)$ is
a short-hand notation for the couple of $i$-$j$ sites.
The terms of ${\mathbb L}$ acting exclusively on the $i$th site can be arbitrary absorbed in the terms of the sum such that $i\in k$.
The observable ${\cal O}$ can be always arranged in a multivariable expansion in powers of $\alpha_k$:
\begin{equation}
  {\cal O} \big( \{\alpha_k\} \big) = \sum_{\{n_k\}}{\cal O}_{\{n_k\}}\prod_k \alpha_k^{n_k}
  \label{M1}
\end{equation}
where $n_k$ runs over all non-negative integers for each $k$,
such that any possible polynomial in the $\alpha_k$ couplings is included.
The expansion~\eqref{M1} can be then reorganized in clusters:
\begin{equation}
  {\cal O} = \sum_{c} W_{[{\cal O}]}(c),
  \label{M2}
\end{equation}
where each $c$ represents a non-empty set of $k$-spatial indexes,
which identify the links belonging to the given cluster.
Specifically, the so called cluster weight $W_{[{\cal O}]}(c)$ contains all terms of the expansion~\eqref{M1}, 
which have at least one power of $\alpha_k, \; \forall k\in c$, and no powers of $\alpha_k$ if $k \notin c$.
Vice-versa, all terms in Eq.~\eqref{M1} can be included in one of these clusters.
Using the inclusion-exclusion principle, one can take $W_{[{\cal O}]}(c)$ out of the sum~\eqref{M2}
obtaining the recurrence relation:
\begin{equation}
  W_{[{\cal O}]}(c) = {\cal O}(c) - \sum_{s \subset c}W_{[{\cal O}]}(s),
  \label{WMc}
\end{equation}
where ${\cal O}(c) = {\rm Tr} \big[ \hat {\cal O} \rho_{\rm SS}(c)\big]$
is the steady-state expectation value of the observable calculated for the finite cluster $c$,
the sum runs over all the subclusters $s$ contained in $c$, and $\rho_{\rm SS}(c)$ is the steady state
of the Liouvillian ${\mathbb L}(c)$ over the cluster $c$.
An important property of Eq.~\eqref{WMc} is that, if $c$ is formed out of two 
disconnected clusters $c_1$ and $c_2$, its weight $W_{[{\cal O}]}(c)$ is zero.
This follows from the fact that ${\cal O}$ is an extensive property (${\cal O}(c) = {\cal O}(c_1) + {\cal O}(c_2)$)
and $c = c_1 + c_2$.

The symmetries of the Liouvillian ${\mathbb L}$ may drastically simplify the summation~\eqref{M2},
since it is typically not needed to compute all the contributions coming from each cluster.
This can be immediately seen, e.g., for situations where the interaction term $\alpha_k$
between different couples of sites is homogeneous throughout the lattice.
In such cases, it is possible to identify the topologically distinct (linked) clusters, so that a representative $c_n$
for each class can be chosen and counted according to its multiplicity $\ell(c_n)$
per lattice site (the lattice constant of the graph $c_n$).
Here the subscript ${}_n$ denotes the number of $k$-spatial indexes that are grouped in the cluster,
that is, its size.
The property ${\cal O}$ per lattice site can be thus written directly in the thermodynamic limit $L \to \infty$ as:
\begin{equation}
  \frac{\cal O}{L} =  \sum_{n=1}^{+\infty} \bigg[ \sum_{\{ c_n \}} \ell(c_n) \, W_{[{\cal O}]}(c_n) \bigg] \,.
  \label{M3}
\end{equation}
The outer sum runs over all possible cluster sizes, while the inner one accounts for all topologically
distinct clusters $\{ c_n \}$ of a given size $n$.
Let us emphasize that, if the series expansion~\eqref{M3} is truncated up to order $n=R$, 
only clusters $c$ at most of size $R$ have to be considered.
Indeed each of them should include at least one power of $\alpha_k, \; \forall k\in c$. 
Therefore a cluster of size $R+1$ or larger does not contribute to the expansion, up to order $\alpha^R$.
As a matter of fact, dealing with open many-body systems significantly reduces our ability to compute large orders
in the expansion, with respect to the {\it closed}-system scenario.
The size of the Liouvillian superoperator governing the dynamics scales as $\dim(\mathbb{L})=d^{2n}$,
where $d$ is the dimension of the local Hilbert space and
$n$ is the number of sites of a given cluster.
In isolated systems, one would need to evaluate the ground state of the cluster Hamiltonian, of size $\dim(\hat H)=d^n$. 
Therefore, for the case of spin-$1/2$ systems ($d=2$), we are able to compute the steady state for clusters up to $n=8$,
such that $\dim(\mathbb{L})=2^{2 \times 8} = 65536$. 
The complexity of the problem is thus comparable to what has been done for spin systems at equilibrium, 
where the NLCE has been computed up to $n=15$ (see, for example, Refs.~\onlinecite{Rigol_2007,Tang_2013}). 

In graph theory, there are established algorithms 
to compute all topologically distinct clusters, for a given size and lattice geometry.
This could drastically increase the efficiency of the NLCE algorithm, since for highly symmetric systems
the number of topologically distinct clusters is exponentially smaller than the total number of connected clusters.
Explaining how to optimize the cluster generation lies beyond the scope of the present work.
The basic cluster generation scheme we used is explained in full detail in Ref.~\onlinecite{Tang_2013}.
Notice that once all the topologically distinct $n$-site clusters and their multiplicities
have been generated for a given lattice geometry, one can employ NLCE for any observable
and Liouvillian within the same spatial symmetry class of the considered lattice. 

A remarkable advantage of NLCE over other numerical methods is that it enables a direct access to the thermodynamic limit,
up to order $R$ in the cluster size, by only counting the cluster contributions of sizes equal or smaller than $R$
(i.e. using a limited amount of resources).
We should stress that, contrary to standard perturbative expansions, there is no perturbative parameter
in the system upon which the NLCE is based and can be controlled.
Properly speaking, the actual control parameter is given by the amount of correlations that are present in the system:
the convergence of the series~\eqref{M3} with $n$ would be ensured from an order $R^\star$
which is larger than the typical length scale of correlations~\cite{Rigol_2006, Tang_2013}. 

In the next sections we give two illustrative examples of how NLCE performs for 2D dissipative
quantum lattice models of interacting spin-1/2 particles.

\section{Model}
\label{sec:Model}

Our model of interest is a spin-$1/2$ lattice system in two dimensions, whose coherent internal dynamics is governed
by the anisotropic XYZ-Heisenberg Hamiltonian:
\begin{equation}
  \hat H = \sum_{\langle i,j\rangle} \left( J_x \hat \sigma_i^x \hat \sigma_j^x 
  + J_y \hat \sigma_i^y \hat \sigma_j^y + J_z \hat \sigma_i^z \hat \sigma_j^z \right) \,,
  \label{eq:Hamiltonian}
\end{equation}
where $\hat \sigma_j^\beta$ ($\beta = x,y,z$) denote the Pauli matrices for the $j$th spin of the system
and $\langle i,j \rangle$ restricts the summation over all couples of nearest neighboring spins.
Each spin is subject to an incoherent dissipative process that tends to
flip it down along the $z$ direction, in an independent way with respect to all the other spins.
In the Markovian approximation, such mechanism is faithfully described by the Lindblad jump operator 
$\hat L_j = \sqrt{\Gamma} \ \hat \sigma^-_j$ acting on each spin:
\begin{equation}
  \mathbb{D}[\rho] = \Gamma \sum_{j} \Big[ \hat \sigma_j^- \rho \, \hat \sigma_j^+
    - \tfrac{1}{2} \big\{ \hat \sigma_j^+ \hat \sigma_j^- , \rho \big\} \Big] \,,
  \label{eq:Lindblad}
\end{equation}
where $\hat \sigma_j^\pm = \frac{1}{2} \left( \hat \sigma_j^x \pm \rmi \, \hat \sigma_j^y \right)$
stands for the corresponding raising and lowering operator along the $z$ axis,
while $\Gamma$ is the rate of the dissipative processes.
in the following we will always work in units of $\Gamma$.

The outlined model is particularly relevant as being considered a prototypical dissipative quantum many-body
system: its phase diagram is very rich and has been subject to a number of studies
at the mean-field level~\cite{Lee_2013} and even beyond such regime,
by means of the cluster mean-field~\cite{Jin_2016}, the corner-space renormalization group~\cite{Rota_2017},
and the dissipative PEPS~\cite{Orus_2016}.
Remarkably, the Lindblad master equation with the Hamiltonian in Eq.~\eqref{eq:Hamiltonian}
and the dissipator in Eq.~\eqref{eq:Lindblad} presents a $\mathbb{Z}_2$ symmetry
which is associated to a $\pi$ rotation along the $z$ axis: $\hat \sigma^x \to -\hat \sigma^x$,
$\hat \sigma^y \to - \hat \sigma^y$.
For certain values of the couplings $J_\alpha$, it is possible to break up this symmetry, thus leading
to a dissipative phase transition from a paramagnetic (PM) to a ferromagnetic (FM) phase,
the order parameter being the in-plane $xy$ magnetization.
We stress that a XY anisotropy ($J_x \neq J_y$) is necessary to counteract the incoherent spin flips,
otherwise the steady-state solution of Eq.~\eqref{eq:Hamiltonian} would be
perfectly polarized, with all the spins pointing down along the $z$ direction.

The existing literature allows us to benchmark our approach, both far from criticality (Sec.~\ref{sec:XXX})
where correlations grow in a controllable way, and in proximity of a $\mathbb{Z}_2$-symmetry breaking
phase transition (Sec.~\ref{sec:XYZ}), where correlations diverge in the thermodynamic limit. 
In the latter we show how it is possible to exploit the NLCE method in combination with a Pad\'e approximants
analysis, in order to calculate the location of the critical point as well as the critical exponent
$\gamma$ of the transition, that is associated to a power-law divergence of the magnetic susceptibility to an external field.
Contrary to all the other known methods, either being mean-field or dealing with finite-length systems,
the NLCE directly addresses the thermodynamic limit and thus, to the best of our knowledge,
at present it represents the only unbiased numerical method to calculate such exponent.

\subsection{Isotropic case}
\label{sec:XXX}

Let us start our analysis by considering a cut in the parameters space which do not cross any critical line.
Specifically we set 
\begin{equation}
\alpha = J_x = -J_y = J_z.
\end{equation}
For $\alpha=0$ the coherent dynamics is switched off, the coupling in $x$-$y$ plane is thus isotropic
and the dissipative processes cannot be counteracted regardless of the value of the local relaxation rates~\cite{Lee_2013}. 
As a consequence, regardless of the initial conditions, the steady-state is the pure state
having all spins pointing down along the $z$-axis:
\begin{equation}
  \label{trivial_ss}
  \left. \rho_{\rm SS} \right|_{\alpha=0}= \bigotimes_i | \!\! \downarrow \rangle \langle \downarrow \!\! | .
\end{equation}
Thus we expect the NLCE would give the exact thermodynamic limit already at first order in the cluster size.
As the parameter $\alpha$ is increased, correlations progressively build up on top of the fully factorizable
density matrix~\eqref{trivial_ss}, therefore higher orders in the expansion of Eq.~\eqref{M3} are needed.

\begin{figure}[t!]
  \centering
  \includegraphics[width=0.9\columnwidth]{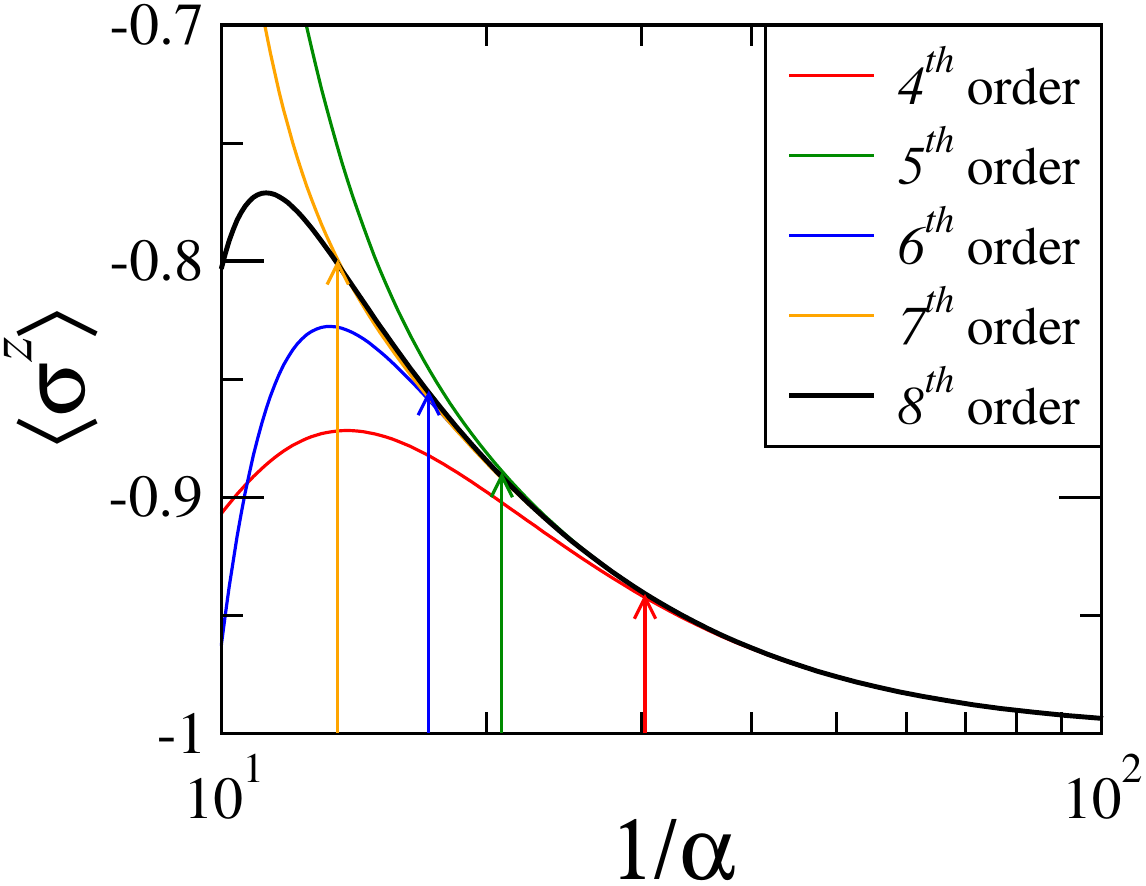}
  \caption{(color online). Steady-state average magnetization along the $z$ direction for the isotropic
    Heisenberg model, evaluated by means of the NLCE (bare sum) at different orders $R$
    in the cluster size, as a function of $1/\alpha$.
    The arrows indicate the values of $\alpha^\star$ at which each curve at the $R$-th order ($R<8$)
    starts deviating significantly from the highest accuracy curve ($R=8$; thick black line)
    that we have.}
  \label{fig:XXX_bare}
\end{figure}

This is exactly what we observe in Fig.\ref{fig:XXX_bare}, where we show the steady-state value
of the average magnetization along the $z$ direction, $\mathcal{O}/L = \langle \hat \sigma^z_j\rangle$,
evaluated by means of the NLCE in Eq~\eqref{M3} up to a given order $R$, as function of $\alpha$.
Note that, as long as $R$ is increased, the convergence of the NLCE to the most accurate
data (highest order that we have) progressively improves.
This shows that, in the region where different curves overlap, correlations among the different sites
are well captured by the clusters that we are considering in the expansion, up to a given order.
When $\alpha$ is increased the range of correlations grows as well, and one needs to
perform the expansion to larger orders.
For $\alpha \gtrsim 0.075$ orders higher than $R=8$ are needed to obtain a good convergence
in the bare data.

\begin{figure}[!b]
  \centering
  \includegraphics[width=0.9\columnwidth]{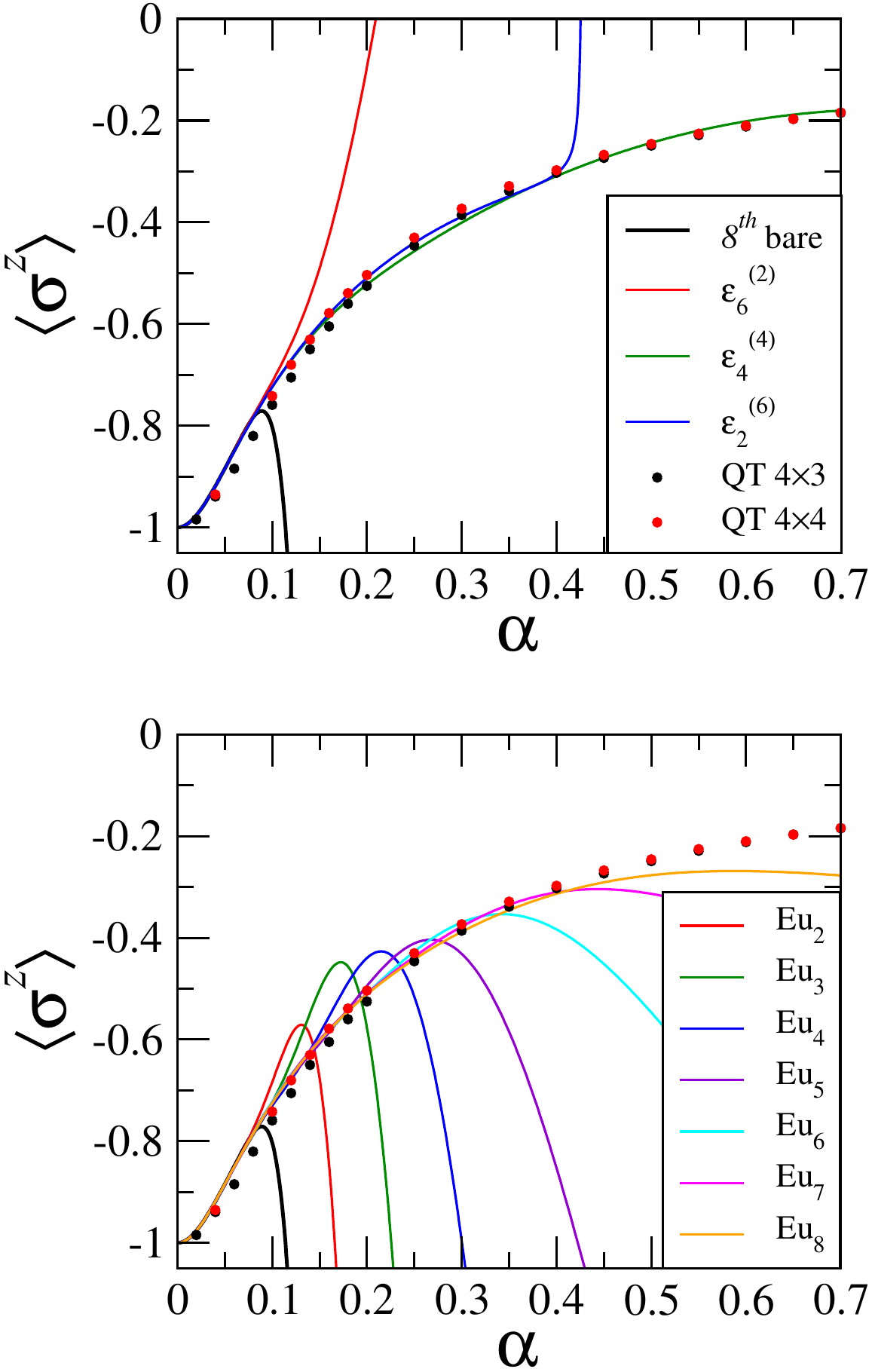}
  \caption{(color online). Steady-state average $z$-magnetization as a function of $\alpha$,
    after implementing two resummation techniques on the bare data at order $R=8$ (black curve,
    same as in Fig.~\ref{fig:XXX_bare}): Wynn's algorithm (colored curves in the upper panel)
    and  Euler transformation (colored curves in the lower panel).
    The symbols denote the results of QT simulations for a finite system using periodic boundary
    conditions, with a $4\times4$ square plaquette (red circles) and a $4\times3$ plaquette (black circles)
    constructed from the previous one after removing the four sites at the corners.}
  \label{fig:XXX_resum}
\end{figure}

It is however possible to improve the convergence of the expansion without increasing
the size of the considered clusters, by simply exploiting two resummation algorithms
that have been already shown to be very useful in the context of NLCEs of given
thermodynamic properties~\cite{Rigol_2006, Tang_2013}.
Specifically we employ the Wynn's algorithm~\cite{wynn_book} and the Euler transformation~\cite{euler_book}.
A detailed explanation on how such resummation schemes can be exploited
in the context of NLCE can be found in Ref.~\onlinecite{Tang_2013}.

The results for $\langle \hat \sigma^z \rangle$ as a function of $\alpha$ are shown in
Fig.~\ref{fig:XXX_resum} for various orders in the two resummation schemes (see legends for details).
It is immediate to see that the convergence of the expansion is drastically improved of about
one order of magnitude.
A comparison of NLCEs data with the outcome of simulations obtained by means of quantum
trajectories (QT)~\cite{Dalibard_1992} for finite-size plaquettes shows that the resummed data
give qualitatively analogous results up to $\alpha \approx 1$, despite a slight discrepancy between them.
Such difference is due to the fact that, even if for small $\alpha$ correlations are very small,
finite-system effects are non-negligible: while NLCEs data are directly obtained in the thermodynamic
limit, QT are inevitably affected by such effects.
As long as $\alpha$ is decreased, the discrepancy between the two approaches decreases,
both leading to $\langle \hat \sigma^z \rangle \to -1$ in the limit $\alpha\to 0$ of Eq.~\eqref{trivial_ss}.

\subsection{Anisotropic case and the paramagnetic to ferromagnetic phase transition}
\label{sec:XYZ}

We now discuss the more interesting scenario of an anisotropic Heisenberg model 
($J_x \neq J_y \neq J_z$), where the system can cross a critical line and exhibit
a dissipative phase transition~\cite{Lee_2013}.
To this purpose, we set
\begin{equation}
  J_x = 0.9, \; J_y = 0.9 + \alpha, \; J_z = 1, \,
  \label{eq:parXYZ}
\end{equation}
with $\alpha \in [0,0.25]$.
For $\alpha=0$ (i.e., $J_x = J_y$), we come back to the trivial situation where the Hamiltonian
conserves the magnetization along the $z$ direction, and the steady state is
the pure state in Eq.~\eqref{trivial_ss}, with all the spins pointing down in the $z$ direction.
Away from this singular point, for a certain $\alpha_c > 0$ the system undergoes
a second-order phase transition associated to the spontaneous breaking of the $\mathbb{Z}_2$
symmetry possessed by the master equation~\eqref{eq:Master}, from a paramagnetic (PM)
for $\alpha < \alpha_c$, to a ferromagnetic (FM) phase for $\alpha > \alpha_c$.
In the FM phase, a finite magnetization in the $x$-$y$ plane develops:
$\langle \hat \sigma^x \rangle, \langle \hat \sigma^y \rangle \neq 0$,
which also defines the order parameter of the transition.

The phenomenology of this phase transition has recently received a lot of attention,
and has been investigated at a Gutzwiller mean-field level~\cite{Lee_2013} and by means
of more sophisticated methods, including the cluster mean-field approach~\cite{Jin_2016},
the corner-space renormalization technique~\cite{Rota_2017}, and
the projected entangled pair operators~\cite{Orus_2016}.
The phase transition point for the same choice of parameters of Eq.~\eqref{eq:parXYZ}
has been estimated to be $\alpha_c = 0.1$~\cite{Lee_2013}, $0.14 \pm 0.01$~\cite{Jin_2016}
and $0.17 \pm 0.02$~\cite{Rota_2017}.

Here we follow the approach of Rota {\it et al.}~\cite{Rota_2017} and discuss the magnetic
linear response to an applied magnetic field in the $x$-$y$ plane,
which modifies the Hamiltonian in Eq.~\eqref{eq:Hamiltonian} according to:
\begin{equation}
  \hat H \to \hat H + \sum_j h \big( \hat \sigma^x_j \cos \theta + \hat \sigma^y_j \sin \theta \big),
  \label{eq:field}
\end{equation}
where $\theta$ denotes the field direction, $\big[ \vec h(\theta) \big] = (h_x, \, h_y)^T$ and
$h_x = h \cos \theta, \; h_y = h \sin \theta$.
Such response is well captured by the susceptibility tensor $\boldsymbol{\chi}$, with matrix
elements $\chi_{\alpha\beta} = \lim_{h_\beta \to 0} \langle \hat \sigma^\alpha \rangle / h_\beta$.
In particular we concentrate on the angularly averaged magnetic susceptibility
\begin{equation}
\label{chiave}
  \chi_{\rm ave} = \lim_{h \to 0} \, \frac{1}{2 \pi} \int_0^{2 \pi} d \theta \, \frac{|\vec M(\theta)|}{h} \,,
\end{equation}
where $\vec M(\theta) = \boldsymbol{\chi} \cdot \vec h(\theta)$ is the induced magnetization along
an arbitrary direction of the field.

\begin{figure}[!t]
  \centering
  \includegraphics[width=0.92\columnwidth]{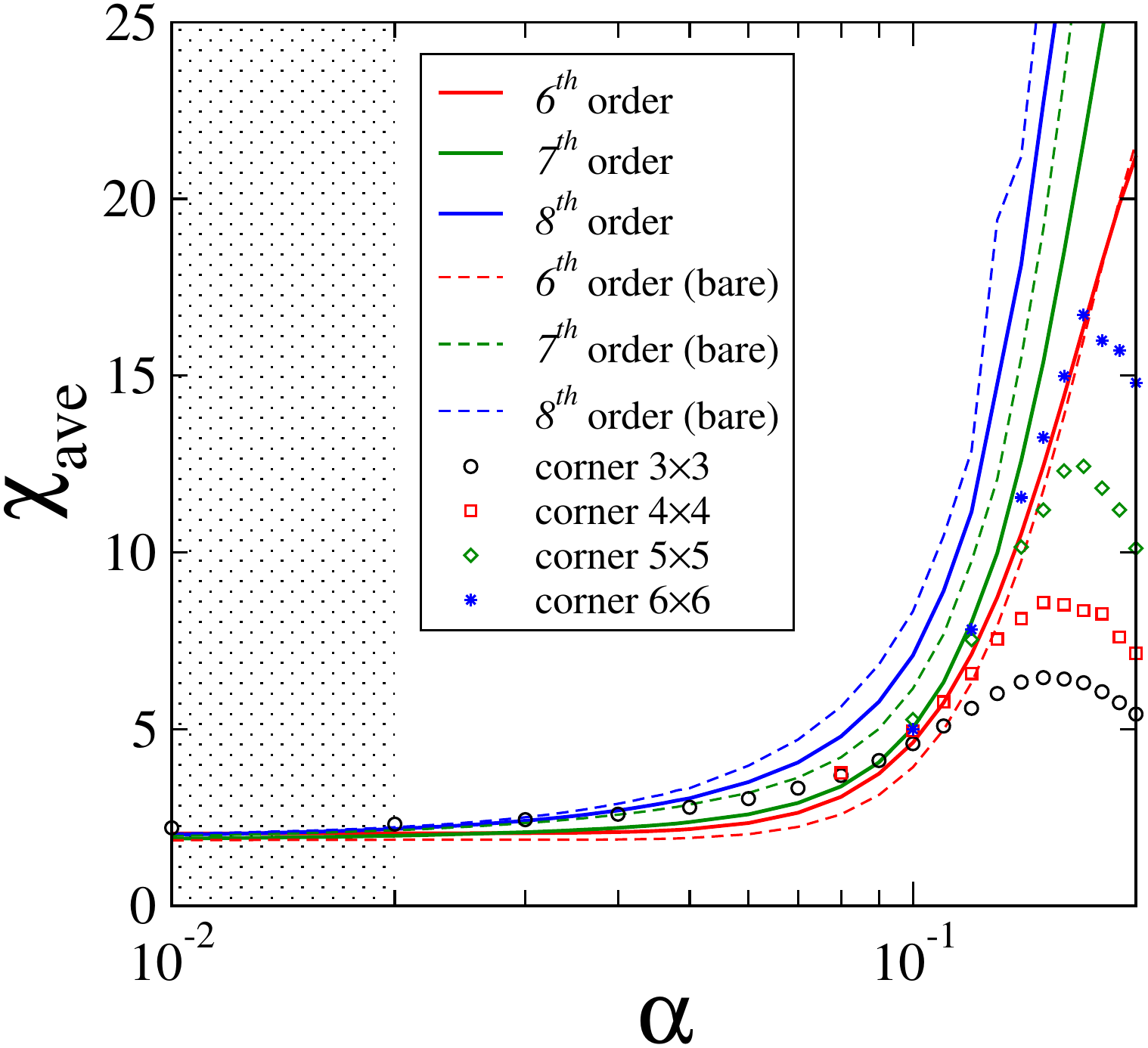}
  \caption{(color online). Angularly averaged magnetic susceptibility to an external field
    in the $x$-$y$ plane, as a function of $\alpha = J_y-0.9$. The continuous curves denote
    Euler resummed data, to the best achievable expansion, of the bare NLCE results up to the order $R=8$.
    The dashed lines are the results of the bare expansions.
    Symbols are the results from the corner-space renormalization method,
    taken from Ref.~\onlinecite{Rota_2017}.
    The dotted area highlights the region of alpha $0\le\alpha\le0.02$, for which the NLCE converges.}
  \label{fig:chiave}
\end{figure}

We start by computing the NLCE for the magnetic susceptibility $\chi_{\rm ave}$
in the parameter range $0 \le \alpha \le 0.25$, and improving the convergence of the series
up to a given order, by exploiting the Euler algorithm.
Along this specific cut in the parameter space, the latter has been proven to be the most
effective (contrary to what we observed far from the criticality -- see Fig.~\ref{fig:XXX_resum}).
The relevant numerical data are shown in Fig.~\ref{fig:chiave}, and are put in direct comparison
with those obtained with an alternative method (the corner-space renormalization group)
in Ref.~\onlinecite{Rota_2017}.
We observe a fairly good agreement with the two approaches, in the small-$\alpha$ parameter range ($0\le\alpha\lesssim0.02$),
and point out that in both cases a sudden increase of $\chi_{\rm ave}$ for $\alpha \gtrsim 0.1$
supports the presence of a phase transition in that region.
It is important to remark that, the result of the expansion at different orders in the uncovered region $\alpha\gtrsim0.02$
has not physical meaning. 
However, as we will show in the next section, by analyzing how the expansion behaves when approaching the criticality, 
it is possible to provide an estimate of the critical point $\alpha_c$, as well as of the critical exponent $\gamma$.
We also note that, contrary to the isotropic case, here we do not observe an {\it exact}
data collapse of the NLCEs for $\chi_{\rm ave}$, even for $\alpha=0$.
The reason resides in the fact that the presence of an external field~\eqref{eq:field} makes
the structure of the steady state nontrivial, as soon as $h \neq 0$, thus admitting
correlations to set in.

\subsubsection{Critical behavior}
\label{sec:critical}

We now show how to exploit the above NLCE data (in combination with a Pad\'e analysis~\cite{Oitmaa_book})
in order to locate the critical point $\alpha_c$ for the PM-FM transition,
and extract the critical exponent $\gamma$ of the magnetic susceptibility~\cite{Sachdev_book}
$\chi_{\rm ave} \sim |\alpha-\alpha_c|^{-\gamma}$.
The possibility to extrapolate the critical exponents for a dissipative quantum phase transition
is very intriguing, since, to the best of our knowledge, the only numerical work in this context,
that is present in the literature, is Ref.~\onlinecite{Rota_2017}.
However, since finite-size systems are considered there, it was only possible to estimate the finite-size
ratio $\gamma/\nu$, where $\nu$ denotes the critical exponent associated to
the divergent behavior of the correlation length.
The present work offers a complementary point of view since here we are able, for the first time,
to provide an independent estimate of the critical exponent $\gamma$ by directly accessing the thermodynamic limit.
 
To achieve this goal we study the logarithmic derivative of the averaged magnetic susceptibility,
which converts an algebraic singularity into a simple pole~\cite{Oitmaa_book}:
\begin{equation}
\label{defdlog}
\Dlog \ \chi_{\rm ave} (\alpha) \equiv \frac{\chi_{\rm ave}' (\alpha)}{\chi_{\rm ave} (\alpha)}.
\end{equation}
If $\chi_{\rm ave} \sim |\alpha-\alpha_c|^{-\gamma}$ for $|\alpha-\alpha_c|\ll1$, the logarithmic derivative behaves as
\begin{equation}
\label{dlog_chiave}
\Dlog \ \chi_{\rm ave} (\alpha)  \sim  \frac{\gamma}{|\alpha-\alpha_c|}.
\end{equation}
Studying the divergent behavior of Eq.~\eqref{dlog_chiave} simplifies the problem,
since the function $\Dlog \ \chi_{\rm ave} (\alpha)$ has a simple pole at the critical point $\alpha=\alpha_c$
with a residue corresponding to the critical exponent $\gamma$.

\begin{figure}[!t]
  \centering
  \includegraphics[width=0.9\columnwidth]{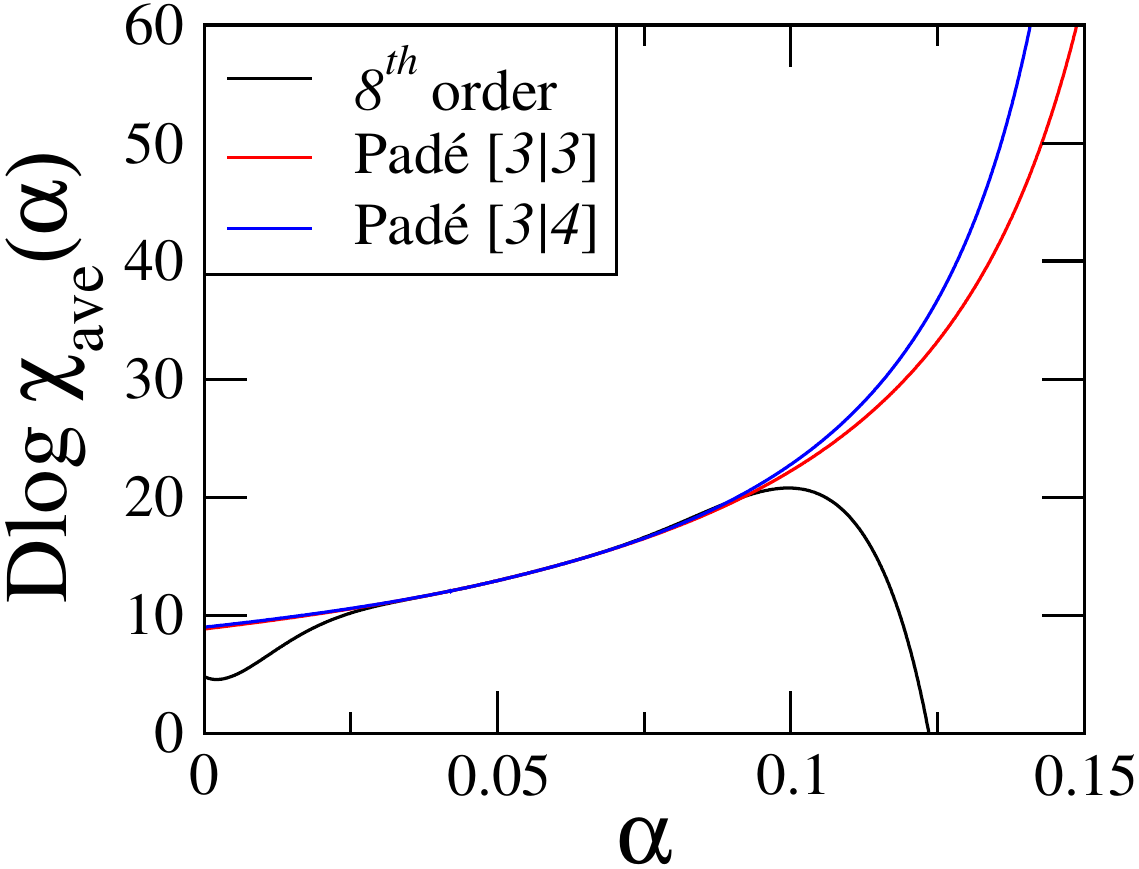}
  \caption{(color online). Logarithmic derivative of $\chi_{\rm ave}$ as a function of $\alpha$.
    The black line is obtained from Euler resummed data to the order $R=8$ (blue line of Fig.~\ref{fig:chiave}).
    The red and blue lines are the results of the Pad\'e analysis with different $[3|3]$ and $[3|4]$
    approximants respectively.}
  \label{fig:XYZ_pade}
\end{figure}

In Fig.~\ref{fig:XYZ_pade} we show the behavior of the logarithmic derivative 
calculated from the Euler resummed data to the order $R=8$ (blue line in Fig.~\ref{fig:chiave}) which represents
our best approximation for $\chi_{ave}$ at small $\alpha$.
The behavior at large $\alpha$ of the function $\Dlog \ \chi_{ave} (\alpha)$ is extrapolated exploiting the Pad\'e approximants.
A Pad\'e approximant is a representation of a finite power series as a ratio of two polynomials
\begin{equation} 
\label{PQpoly}
\Dlog \ \chi_{ave} (\alpha) = \sum_{n=0}^R a_n\alpha^n = \frac{P_L(\alpha)}{Q_M(\alpha)},
\end{equation}
where $P_L(\alpha)$ and $Q_M(\alpha)$ are polynomials of degree $L$ and $M$ (with $L+M\le R$), respectively.
This is denoted as the $[L|M]$ approximant, and can represent functions with simple poles {\it exactly}.
Next, we fit $\Dlog \ \chi_{\rm ave}(\alpha)$ (black line in Fig.~\ref{fig:XYZ_pade}) with an $8$-th degree polynomial
between $\alpha_{\rm in}=0.05$ to $0.06\le\alpha_{\rm fin}\le0.1$ in order to obtain
the coefficients $\{a_n\}_{n=1,\dots,R}$ (with $R=8$).
Once the coefficients $\{a_n\}$ are known, it is straightforward to evaluate the coefficients of the polynomials
$P_L$ and $Q_M$ through Eq.~\eqref{PQpoly}.
Further details about this procedure can be found in App.~\ref{app:pade}.
As is clear from Eq.~\eqref{PQpoly}, the position of the critical point $\alpha_c$ can be deduced by studying the zeroes of $Q_M(\alpha)$. 
Typically, only one of the $M$ zeros is real and located in the region of interest.
Finally, the critical exponent is evaluated by computing the residue of $Q_M(\alpha)$ at $\alpha=\alpha_c$:
\begin{equation}
\gamma = - \lim_{\alpha\to\alpha_c} Q_M(\alpha) (\alpha-\alpha_c).
\end{equation}

Of course, the values of $\alpha_c$ and $\gamma$ will depend on the specific choice of the approximates $[L,M]$ and on the region
over which the fit is performed.
The dependence of the results on $\alpha_{\rm fin}$ is shown in App.~\ref{app:pade}.
We found that the Pad\'e analysis gives stable results for $0.06\lesssim\alpha_{\rm fin}\lesssim0.08$ and $0.06\lesssim\alpha_{\rm fin}\lesssim0.095$
for the approximants $[3|3]$ and $[3|4]$ respectively.

The results of the Pad\'e analysis hint for a divergence at $\alpha_c=0.179\pm0.001$ with $\gamma=1.85\pm0.05$ for $[3|3]$,
and $\alpha_c=0.1665\pm0.0005$ with $\gamma=1.5\pm0.05$ for [3|4]. 
The other approximants $[L|M]$ such that $L+M\le R=8$ do not give physical results in this range of parameters.
The error bar is underestimated, since it accounts only for the error introduced in the fitting procedure and neglects the propagation of the numerical error made on the steady-state evaluation.
Furthermore, the Pad\'e analysis has been performed over a range of $\alpha$ for which the resummed NLCE is not exactly converged (see Fig.~\ref{fig:chiave}). 
To overcome this issue, one should be able to compute higher orders in the expansion and to perform a more accurate analysis of the criticality.
However, the value of the critical point we found is in agreement with the results reported
in Ref.~\onlinecite{Jin_2016} and Ref.~\onlinecite{Rota_2017}, so far.

\section{Conclusions}
\label{sec:concl}

In this work we have proposed a numerical algorithm based on the generalization of the linked-cluster expansion
to open quantum systems on a lattice, allowing to directly access the thermodynamic limit
and to evaluate extensive properties of the system.
Specifically, we extended the formalism to the Liouvillian case and showed how the basic properties
of the expansion are translated to the open-system realm.
Given its generality, this method can be applied to open fermionic, bosonic and spin systems
in an arbitrary lattice geometry.

We tested our approach with a study of the steady-state properties of the paradigmatic dissipative spin-1/2 XYZ model
on a two-dimensional square lattice.
Far away from the critical boundaries of the model, we accurately computed the spin magnetization.
Upon increasing the order of the expansion, we were able to progressively access regions of the phase diagram
that are characterized by a larger amount of correlations among distant sites.  
The convergence properties of the expansion can be dramatically improved by employing more sophisticated resummation schemes.
We then used the numerical linked-cluster expansion across a phase transition in order to study its critical properties.
By means of a Pad\'e analysis of the series, we located the critical point and provided the first estimate of the 
critical exponent $\gamma$, which determines the divergent behavior of the (average) magnetic susceptibility
close to the phase transition.

At present, this method together with the one in Ref.~\onlinecite{Orus_2016}
are the only (non mean-field) numerical approaches that allow to compute
the steady-state properties of an open lattice model in two spatial dimensions in the thermodynamic limit.
Here the intrinsic limitation is that, in order to compute high-order terms in the expansion
(and thus to access strongly correlated regions of the phase space),
the evaluation of the steady state on a large number of connected sites is required.
Furthermore, in the case of bosonic systems, a further complication arises from the local Hilbert space dimension.
We believe that a very interesting perspective left for the future, is the combination of the linked-cluster expansion
with the corner-space renormalization method~\cite{Rota_2017}, and also possibly with Monte Carlo approaches~\cite{savona_private}.   
Additionally, a careful identification of the internal symmetries of the model may help in decreasing the effective dimension of the Liouvillian space.

\acknowledgments

We thank M. C\`e, L. Mazza, and R. Rota for fruitful discussions.
We acknowledge the CINECA award under the ISCRA initiative, for the availability of high performance computing resources and support.
AB and CC acknowledge support from ERC (via Consolidator Grant CORPHO No.~616233). 
RF acknowledges support by EU-QUIC, CRF, Singapore Ministry of Education, CPR-QSYNC, SNS-Fondi interni 2014, and the Oxford Martin School.
JJ acknowledges support from the National Natural Science Foundation of China No.~11605022,
Natural Science Foundation of Liaoning Province No.~2015020110, and the Xinghai Scholar Cultivation Plan
and the Fundamental Research Funds for the Central Universities.
OV thanks Fundaci\'on Rafael del Pino, Fundaci\'on Ram\'on Areces and RCC Harvard.

\appendix

\section{Pad\'e approximants}
\label{app:pade} 

Here we discuss the details related to the Pad\'e analysis of the divergent behavior of the magnetic susceptibility, which has been performed in Sec.~\ref{sec:critical}.
As already introduced in the main text, the Pad\'e approximant is a representation of the first $R$ terms of a power series as a ratio of two polynomials.

Let us consider Eq.~\eqref{PQpoly}, where $\Dlog \chi_{\rm ave}(\alpha)$ is the function for which we know the Taylor expansion up to the order $R$
\begin{equation}
\label{coef_f}
\Dlog \chi_{\rm ave}(\alpha) = \sum_{n=0}^{R} a_n \alpha^n = \frac{P_L(\alpha)}{Q_M(\alpha)}, 
\end{equation}
and the Pad\'e polynomials are parametrised as follow
\begin{equation}
\label{coef_pq}
P_L(\alpha) = \sum_{n=0}^L p_n \alpha^n,  \quad Q_M(\alpha) = 1+ \sum_{n=1}^M q_n \alpha^n ,
\end{equation}
with $L+M\le R$.
This is denoted as the $[L|M]$ approximant.

Let us start by showing that if the function $\chi_{\rm ave}(\alpha)$ has an algebraic singularity at $\alpha=\alpha_c$ then its logarithmic derivative (see Eq.\eqref{defdlog}) has a simple pole at the same value of $\alpha$.
To show this, let us note that for $|\alpha-\alpha_c|\ll1$ 
\begin{equation}
\label{chiave_app}
\chi_{\rm ave}(\alpha) =\frac{g(\alpha)}{|\alpha-\alpha_c|^{\gamma}},
\end{equation}
where $g(\alpha)$ is an analytic function in the range of $\alpha$ we are interested in.
So that Eq.~\eqref{chiave_app} becomes
\begin{equation}
\Dlog \ \chi_{\rm ave}(\alpha) = \frac{g'(\alpha)}{g(\alpha)} - \frac{\gamma}{|\alpha-\alpha_c|}.
\end{equation}

\begin{figure}[!t]
  \centering
  \includegraphics[width=0.9\columnwidth]{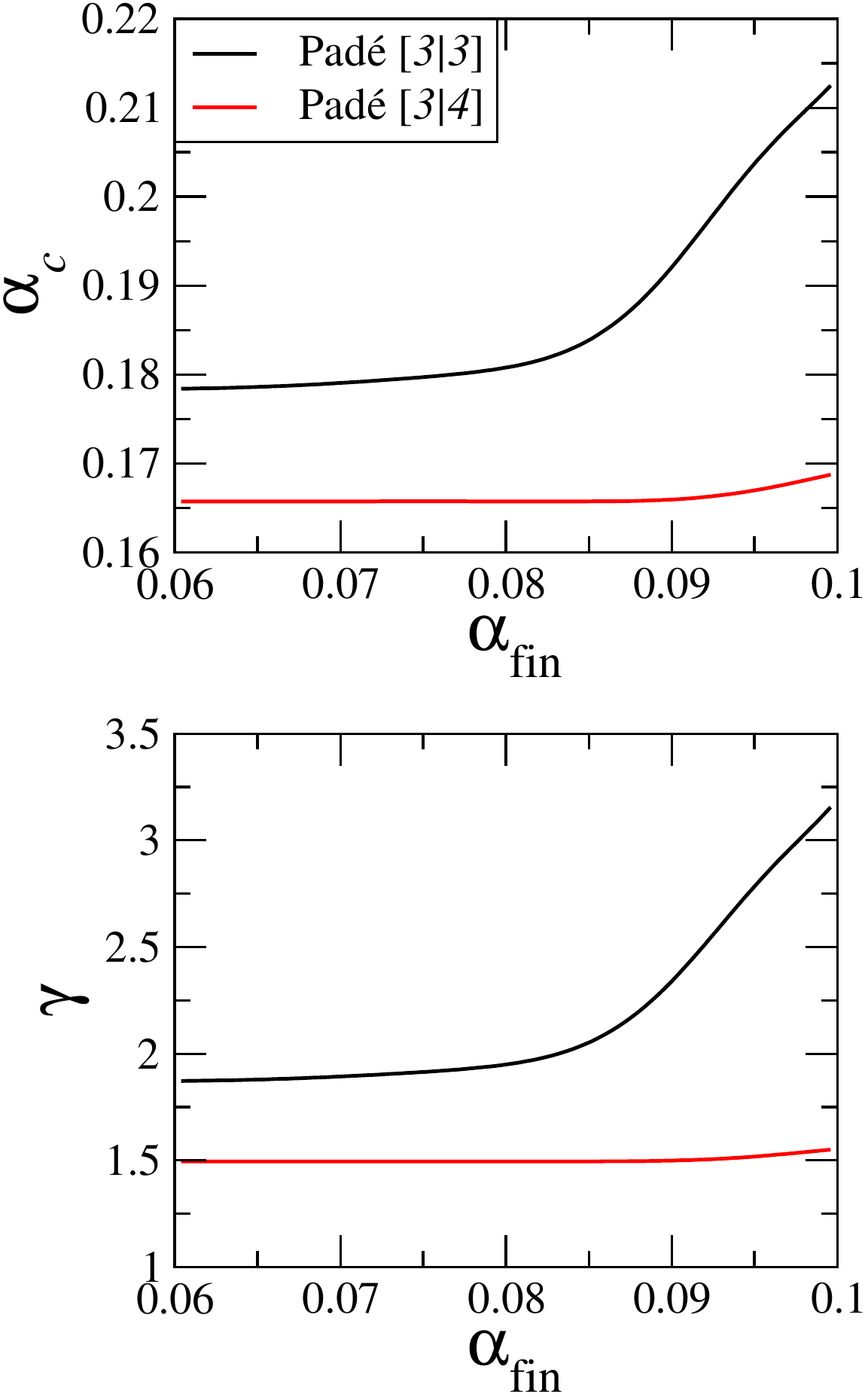}
  \caption{(color online). Position of the critical point $\alpha_c$ (top panel) and value of the critical exponent $\gamma$ (bottom panel)
    as a function of the upper boundary of the fitting region $\alpha_{\rm fin}$}
  \label{fig:XYZ_pade_stab}
\end{figure}

Given the coefficients $\{a_n\}$ (calculated by fitting the function $\Dlog \chi_{\rm ave}(\alpha)$ with an $8$-th degree polynomial from $\alpha_{\rm in}\le\alpha\le\alpha_{\rm fin}$), it is easy to obtain the coefficients $\{p_n\}$ and $\{q_n\}$ in Eq.~\eqref{coef_pq} exploiting Eq.~\eqref{coef_f}.
This gives the following set of $L+M+1$ linear equations
\begin{eqnarray}
\label{coef_sist}
a_0 &=& p_0 \cr
&&\cr
a_1+a_0q_1&=&p_1\cr
&&\cr
a_2+a_1q_1+a_0q_2&=&p_2\cr
\vdots\cr
a_L+a_{L-1}q_1+\cdots+a_0q_L&=&p_L\cr
&&\cr
a_{L+1}+a_Lq1+\cdots+a_{L-M+1}q_M&=&0 \cr
\vdots\cr
a_{L+M}+a_{L+M-1}q1+\cdots+a_Lq_M &=&0.
\end{eqnarray}

Once the coefficients $\{p_n\}$ and $\{q_n\}$ has been determined, one can calculate $\alpha_c$ by studying the zeroes of $Q_M(\alpha)$ and compute the critical exponent $\gamma$ by evaluating the residue at $\alpha=\alpha_c$ (see Sec.~\ref{sec:critical}).
In Fig.~\ref{fig:XYZ_pade_stab} we show the position of the critical point $\alpha_c$ (top panel) and the value of the critical exponent $\gamma$ (bottom panel) as a function of the upper fit boundary $\alpha_{\rm fin}$ for $\alpha_{\rm in}=0.05$.


\begin{thebibliography}{100}

\bibitem{kasprzak2006}
  J. Kasprzak, M. Richard, S. Kundermann, A. Baas, P. Jeambrun, J. M. J. Keeling, F. M. Marchetti, 
  M. H. Szymanska, R. Andr\'e, J. L. Staehli, V. Savona, P. B. Littlewood, B. Deveaud, and Le Si Dang, Nature {\bf 443}, 409 (2006).
  
\bibitem{syassen2008}
  N. Syassen, D. M. Bauer, M. Lettner, T. Volz, D. Dietze, J. J. Garc\'ia-Ripoll, J. I. Cirac, G. Rempe, and S. D\"urr, Science {\bf 320}, 1329 (2008).

\bibitem{baumann2010} 
  K. Baumann, C. Guerlin, F. Brennecke, and T. Esslinger,
  Nature {\bf 464}, 1301 (2010).

\bibitem{Muller_2012}
  M. M\"uller, S. Diehl, G. Pupillo, and P. Zoller,  Adv. At. Mol. Opt. Phys. {\bf 61}, 1 (2012).

\bibitem{Houck_2012}
  A. A. Houck, H. E. T\"ureci, and J. Koch,  Nat. Phys. {\bf 8}, 292 (2012).

\bibitem{Ludwig_2013}
  M. Ludwig and F. Marquardt,  Phys. Rev. Lett. {\bf 111}, 073603 (2013).

\bibitem{Tomadin_rev}
  A. Tomadin and R. Fazio,  J. Opt. Soc. Am. B {\bf 27}, A130 (2010).

\bibitem{Hartmann_rev}
  M. Hartmann,  J. Opt. {\bf 18}, 104005 (2016).

\bibitem{Sieberer_2016}
  L. M. Sieberer, M. Buchhold, and S. Diehl,  Rep. Prog. Phys. {\bf 79}, 096001 (2016).

\bibitem{LeHur_rev}
  K. Le Hur, L. Henriet, A. Petrescu, K. Plekhanov, G. Roux, and M. Schir\'{o},  C. R. Physique {\bf 17}, 808 (2016).

\bibitem{Angelakis_rev}
  C. Noh and D. Angelakis,  Rep. Prog. Phys. {\bf 80}, 016401 (2017).

\bibitem{Tomadin2010}
  A. Tomadin, V. Giovannetti, R. Fazio, D. Gerace, I. Carusotto, H. E. T\"ureci, and A. Imamoglu,
  Phys. Rev. A {\bf 81}, 061801 (2010). 

\bibitem{Carusotto_2009}
  I. Carusotto, D. Gerace, H. E. T\"ureci, S. De Liberato, C. Ciuti, and A. Imamoglu,
  Phys. Rev. Lett. {\bf 103}, 033601 (2009).

\bibitem{Grujic_2012}
  T. Grujic, S. R. Clark, D. G. Angelakis, and D. Jaksch,  New J. Phys. {\bf 14}, 103025 (2012);
  T. Grujic, S. R. Clark, D. Jaksch, and D. G. Angelakis,  Phys. Rev. A {\bf 87}, 053846 (2013).

\bibitem{Rivas_2014}
  J. Ruiz-Rivas, E. del Valle, C. Gies, P. Gartner, and M. J. Hartmann, Phys. Rev. A {\bf 90}, 033808 (2014).

\bibitem{biella2015}
  A. Biella, L. Mazza, I. Carusotto, D. Rossini, and R. Fazio,  Phys. Rev. A {\bf 91}, 053815 (2015).  

\bibitem{angelakis2015}
  C. Lee, C. Noh, N. Schetakis, and D. G. Angelakis,  Phys. Rev. A {\bf 92}, 063817 (2015).  

\bibitem{mertz2016}
  T. Mertz, I. Vasi\'c, M. J. Hartmann, and W. Hofstetter,  Phys. Rev. A {\bf 94}, 013809 (2016). 

\bibitem{savona2017_01}
  K. Debnath, E. Mascarenhas, V Savona,  arXiv:1706.04936 (2017).

\bibitem{savona2017_02}
  J. Reisons, E. Mascarenhas, V. Savona,  Phys. Rev. B {\bf 96}, 165137 (2017).
  
\bibitem{Diehl2008}
  S. Diehl, A. Micheli, A. Kantian, B. Kraus, H. P. B\"uchler, and P. Zoller, Nat. Phys. {\bf 4}, 878 (2008).

\bibitem{Verstraete2009}
  F. Verstraete, M. M. Wolf, and J. I. Cirac, Nat. Phys. {\bf 5}, 633 (2009).  
  
\bibitem{hartmann2010}
  M. J. Hartmann,  Phys. Rev. Lett. {\bf 104}, 113601 (2010).

\bibitem{umucalilar2012}
  R. O. Umucalilar and I. Carusotto,  Phys. Rev. Lett. {\bf 108}, 206809 (2012).  
  
\bibitem{jin2013}
  J. Jin, D. Rossini, R. Fazio, M. Leib, and M. J. Hartmann,  Phys. Rev. Lett. {\bf 110}, 163605 (2013).
  J. Jin, D. Rossini, M. Leib, M. J. Hartmann, and R. Fazio,  Phys. Rev. A {\bf 90}, 023827 (2014).
  
\bibitem{Yuge_2014}
  T. Yuge, K. Kamide, M. Yamaguchi, and T. Ogawa, J. Phys. Soc. Jpn. {\bf 83}, 123001 (2014).
  
\bibitem{hoening2014}
  M. Hoening, W. Abdussalam, M. Fleischhauer, and T. Pohl,  Phys. Rev. A {\bf 90}, 021603(R) (2014).
  
\bibitem{chan2015}	
  C.-K. Chan, T. E. Lee, and S. Gopalakrishnan,  Phys. Rev. A {\bf 91}, 051601 (2015).
  
\bibitem{wilson2016}
  R. M. Wilson, K. W. Mahmud, A. Hu, A. V. Gorshkov, M. Hafezi, and M. Foss-Feig,  Phys. Rev. A {\bf 94}, 033801 (2016).

\bibitem{ff2017}
  M. Foss-Feig, P. Niroula, J. T. Young, M. Hafezi, A. V. Gorshkov, R. M. Wilson, and M. F. Maghrebi,
  Phys. Rev. A {\bf 95}, 043826 (2017).
  
\bibitem{Lee_2011}
  T. E. Lee, H. H\"affner, and M. C. Cross,  Phys. Rev. A {\bf 84}, 031402 (2011).  
  
\bibitem{Lee_2013}
  T. E. Lee, S. Gopalakrishnan, and M. D. Lukin,  Phys. Rev. Lett. {\bf 110}, 257204 (2013).   
  
\bibitem{savona2017}
  V. Savona, Phys. Rev. A {\bf 96}, 033826 (2017).
  
\bibitem{jose2017}
  J. Lebreuilly, A. Biella, F. Storme, D. Rossini, R. Fazio, C. Ciuti, I. Carusotto,  Phys. Rev. A {\bf 96}, 033828 (2017).

\bibitem{biella2017}
  A. Biella, F. Storme, J. Lebreuilly, D. Rossini, R. Fazio, I. Carusotto, C. Ciuti,  Phys. Rev. A {\bf 96}, 023839 (2017).

\bibitem{torre2012}
  E. G. Dalla Torre, E. Demler, T. Giamarchi, and E. Altman,  Phys. Rev. B {\bf 85}, 184302 (2012).

\bibitem{sieberer2013}
  L. M. Sieberer, S. D. Huber, E. Altman, and S. Diehl,  Phys. Rev. Lett. {\bf 110}, 195301 (2013).

\bibitem{marino2016}
  J. Marino and S. Diehl,  Phys. Rev. Lett. {\bf 116}, 070407 (2016).
  
\bibitem{Rota_2017}
  R. Rota, F. Storme, N. Bartolo, R. Fazio, and C. Ciuti,  Phys. Rev. B {\bf 95}, 134431 (2017).

\bibitem{Fitzpatrick_2017}
  M. Fitzpatrick, N. M. Sundaresan, A. C. Y. Li, J. Koch, and A. A. Houck,  Phys. Rev. X {\bf 7}, 011016 (2017).

\bibitem{Petruccione_book}
  H.-P. Breuer and F. Petruccione, {\it The Theory of Open Quantum Systems} (Oxford University Press, New York, 2002).

\bibitem{Rivas_book}
  A. Rivas and S. F. Huelga, \textit{Open Quantum Systems. An Introduction} (Springer, Heidelberg, 2011).
  
\bibitem{albert2014}  
  V. V. Albert and L. Jiang, Phys. Rev. A {\bf 89}, 022118 (2014).
  
\bibitem{Prosen_2008}
  T. Prosen,  New J. Phys. {\bf 10}, 043026 (2008).

\bibitem{Li_2016}
  A. C. Y. Li, F. Petruccione, and J. Koch,  Sci. Rep. {\bf 4}, 4887 (2014);  Phys. Rev. X {\bf 6}, 021037 (2016).

\bibitem{Maghrebi2015}
  M. F. Maghrebi and A. V. Gorshkov, Phys. Rev. B {\bf 93}, 014307 (2016).  

\bibitem{Verstraete_2004}
  F. Verstraete, J. J. Garc\'ia-Ripoll, and J. I. Cirac,  Phys. Rev. Lett. {\bf 93}, 207204 (2004).

\bibitem{Zwolak_2004}
  M. Zwolak and G. Vidal,  Phys. Rev. Lett. {\bf 93}, 207205 (2004).

\bibitem{Prosen_2009}
  T. Prosen and M. Znidaric,  J. Stat. Mech. (2009) P02035.
  
\bibitem{Cui_2015}
  J. Cui, J. I. Cirac, and M. C. Ba\~nuls,  Phys. Rev. Lett. {\bf 114}, 220601 (2015).

\bibitem{Mascarenhas_2015}
  E. Mascarenhas, H. Flayac, and V. Savona,  Phys. Rev. A {\bf 92}, 022116 (2015).

\bibitem{Werner_2016}
  A. H. Werner, D. Jaschke, P. Silvi, M. Kliesch, T. Calarco, J. Eisert, and S. Montangero,  Phys. Rev. Lett. {\bf 116}, 237201 (2016).
  
\bibitem{Jin_2016}
  J. Jin, A. Biella, O Viyuela, L. Mazza, J. Keeling, R. Fazio, and D. Rossini,  Phys. Rev. X {\bf 6}, 031011 (2016).

\bibitem{Degenfeld_2014}
  P. Degenfeld-Schonburg and M. J. Hartmann,  Phys. Rev. B {\bf 89}, 245108 (2014).

\bibitem{Weimer_2015}
  H. Weimer,  Phys. Rev. Lett. {\bf 114}, 040402 (2015).

\bibitem{Casteels_2016}
  W. Casteels, S. Finazzi, A. Le Boit\'e, F. Storme, and C. Ciuti,  New J. Phys. {\bf 18}, 093007 (2016).

\bibitem{Finazzi_2015}
  S. Finazzi, A. Le Boit\'e, F. Storme, A. Baksic, and C. Ciuti,  Phys. Rev. Lett. {\bf 115}, 080604 (2015).
  
\bibitem{Orus_2016}
  A. Kshetrimayum, H. Weimer, and R. Orus, Nat. Commun. {\bf 8}, 1291 (2017).

\bibitem{tsuji2009}
  N. Tsuji, T. Oka, and H. Aoki, Phys. Rev. Lett. {\bf 103}, 047403 (2009).

\bibitem{amaricci2012}
  A. Amaricci, C. Weber, M. Capone, and G. Kotliar, Phys. Rev. B {\bf 86}, 085110 (2012).

\bibitem{aoki2014}
  H. Aoki, N. Tsuji, M. Eckstein, M. Kollar, T. Oka, and P. Werner, Rev. Mod. Phys. {\bf 86}, 779 (2014).

\bibitem{Oitmaa_book}
  J. Oitmaa, C. Hamer, and W. Zheng, {\it Series expansion methods for strongly interacting lattice models},
  (Cambridge University Press, Cambridge, 2006).

\bibitem{Yang_1952}
  C. N. Yang and T. D. Lee,  Phys. Rev. {\bf 87}, 404 (1952);  {\it ibid.} {\bf 87}, 410 (1952).

\bibitem{Rigol_2006}
  M. Rigol, T. Bryant, and R. R. P. Singh,  Phys. Rev. Lett. {\bf 97}, 187202 (2006).

\bibitem{Rigol_2007}
  M. Rigol, T. Bryant, and R. R. P. Singh,  Phys. Rev. E {\bf 75}, 061118 (2007);  {\it ibid.} {\bf 75}, 061119 (2007).
  
\bibitem{Rigol_2014}
  M. Rigol,  Phys. Rev. Lett. {\bf 112}, 170601 (2014).

\bibitem{Mallayya_2017}
  K. Mallayya and M. Rigol,  Phys. Rev. E {\bf 95}, 033302 (2017).

\bibitem{Bruognolo_2017}
  B. Bruognolo, Z, Zhu, S. R. White, and E. M. Stoudenmire,  arXiv:1705.05578 (2017).
  
\bibitem{buca2017}
B. Bu{\v c}a, and T. Prosen, arXiv:1710.08319 (2017).

\bibitem{biondi2015}
M. Biondi, E. P. L. van Nieuwenburg, G. Blatter, S. D. Huber, and S. Schmidt, Phys. Rev. Lett. {\bf 115}, 143601 (2015).

\bibitem{Tang_2013}
  B. Tang, E. Khatami, and M. Rigol,  Comp. Phys. Commun. {\bf 184}, 557 (2013).
  
\bibitem{wynn_book}
  A. J. Guttmann, {\it Phase Transitions and Critical Phenomena}, Vol.13 (Academic Press, London, 1989).
  
\bibitem{euler_book}
  H. W. Press, B. P. Flannery, S. A. Teukolsky, and W. T. Vetterling, {\it Numerical Recipes in Fortran} (Cambridge University Press, Cambridge, England, 1999).

\bibitem{Dalibard_1992}
  J. Dalibard, Y. Castin, and K. M\o lmer,  Phys. Rev. Lett. {\bf 68}, 580 (1992).

\bibitem{Sachdev_book}
  S. Sachdev, {\it Quantum Phase Transitions} (Cambridge University Press, Cambridge, England, 2000).
  
\bibitem{savona_private}
  A. Nagy and V. Savona, private communication.
  
\bibitem{euler_book2}
  K. Knopp, {\it Theory and Application of Infinite Series} (Dover Publications, New York, 1990).  

\end{thebibliography}
\end{document}